# Unsupervised String Transformation Learning for Entity Consolidation

Dong Deng[†] Wenbo Tao[†] Ziawasch Abedjan[◇] Ahmed Elmagarmid[♣] Guoliang Li[‡]
Ihab F. Ilyas[♡] Samuel Madden[†] Mourad Ouzzani[♣] Michael Stonebraker[†] Nan Tang[♣]

[†]MIT [◇]TU Berlin [♣]Qatar Computing Research Institute, HBKU [‡]Tsinghua University [♡]University of Waterloo
{dongdeng, wenbo, stonebraker, madden}@csail.mit.edu, abedjan@tu-berlin.de
{aelmagarmid, mouzzani, ntang}@hbku.edu.qa, liguoliang@tsinghua.edu.cn, ilyas@uwaterloo.ca

## ABSTRACT

Data integration has been a long-standing challenge in data management with many applications. A key step in data integration is entity consolidation. It takes a collection of clusters of duplicate records as input and produces a single "golden record" for each cluster, which contains the canonical value for each attribute. Truth discovery and data fusion methods as well as Master Data Management (MDM) systems can be used for entity consolidation. However, to achieve better results, the variant values (i.e., values that are logically the same with different formats) in the clusters need to be consolidated before applying these methods.

For this purpose, we propose a data-driven method to standardize the variant values based on two observations: (1) the variant values usually can be transformed to the same representation (e.g., "Mary Lee" and "Lee, Mary") and (2) the same transformation often appears repeatedly across different clusters (e.g., transpose the first and last name). Our approach first uses an unsupervised method to generate groups of value pairs that can be transformed in the same way (i.e., they share a transformation). Then the groups are presented to a human for verification and the approved ones are used to standardize the data. In a real-world dataset with 17,497 records, our method achieved 75% recall and 99.5% precision in standardizing variant values by asking a human 100 yes/no questions, which completely outperformed a state of the art data wrangling tool.

## 1 INTRODUCTION

Data integration plays an important role in many real-world applications such as customer management, supplier management, direct marketing, and comparison shopping. Two key steps in data integration are entity resolution and entity consolidation. Entity resolution [18] produces clusters of records thought to represent the same entity from disparate data sources. Many resolution methods [6, 12, 40, 42] and systems have been developed in recent years, such as Tamr [38], Magellan [30], and DataCivilizer [13].

Entity consolidation takes as input a collection of clusters, where each cluster contains a set of duplicate records, and outputs a single "golden record" for each cluster, which represents the canonical value for each attribute in the cluster. As the attribute values of two duplicate records may not necessarily be redundant, we cannot simply choose an arbitrary record from each cluster as the golden record. For example, $r_4[Address]$ ="5 th St, 22701 California" and $r_5[Address]$ ="3rd E Ave, 33990 California" in two duplicate records in Table 1 refer to different addresses and thus are not redundant. Instead, they conflict with each other.

Probabilistic methods have been proposed to resolve conflicts in truth discovery and data fusion [15, 17, 44]. They can be adapted for entity consolidation. Master Data Management (MDM) systems leverage human-written rules for entity consolidation. However, to achieve better results, the variant values (values that are logically the same with different formats) in the same clusters should be consolidated before applying existing methods. For this purpose, in this paper, we propose a data-driven method to standardize the variant values. As an example, as shown in Figure 1, our method takes Table 1 as input and outputs Table 2. Afterwards, existing entity consolidation methods can take Table 2 as input and construct the golden records in Table 3.

**Solution Overview.** We propose a data-driven approach to identify and standardize the variant values in clusters. Because the variant values usually can be transformed to human is satisfied with the result. The reason for this is twofold. First, larger groups are more 'profitable' once they are approved by the human. Second, larger groups are more likely to be correct as the variant values often share common transformations that appear repeatedly across different clusters (e.g., both "Mary Lee"↔"Lee, Mary" and "James Smith"↔"Smith, James" can be transformed by transposing the first and last name). Finally, the approved groups will be used to perform the replacement and update the clusters.

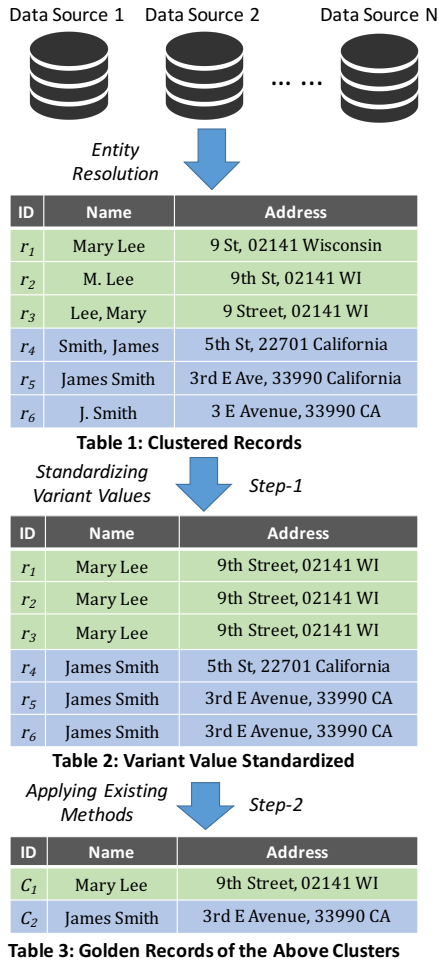

**Figure 1: An example**

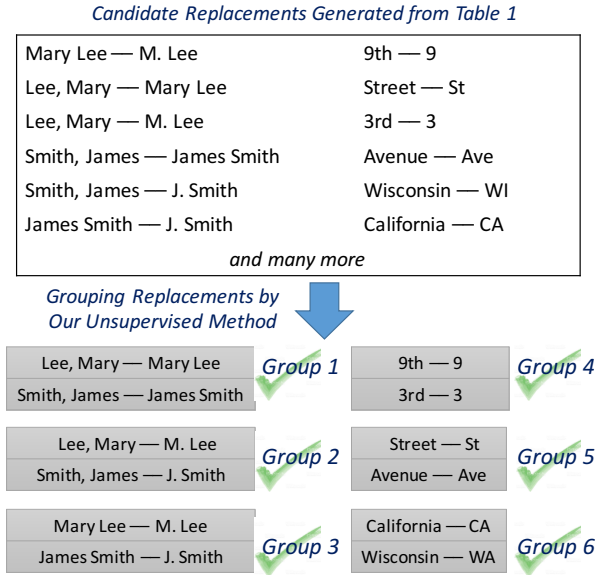

**Figure 2: Grouping candidate replacements**

**Unsupervised Group Generation.** Clearly, to save human effort, it is desirable for the number of groups to be as small as possible. Thus our goal is to group all the value pairs such that the value pairs within the same group can be transformed in the same way (i.e., share the same transformation), and the number of groups is minimized. To formally express the transformation (which describes how one string is transformed to another), we borrow and extend a powerful domain specific language (DSL) designed by Guwani et al [22, 23]. The DSL is very expressive and has been used in production in Microsoft Excel. However, using this DSL, each value pair can be transformed in an exponential number of ways. Moreover, we can prove it is NP-complete to optimally group the value pairs based on our criteria.

To alleviate this problem, we develop a simple and effective, data-driven greedy strategy, along with optimization techniques, to group the value pairs. However, this approach incurs a large upfront time cost as it generates all the groups at once. To address this issue, we design an incremental (i.e., top-k) algorithm which generates the next largest group with each invocation. It reduces the upfront cost by up to 3 orders of magaintitude in our experiments. We compared with a baseline method that uses the data wrangling tool

---

each other (e.g., "Mary Lee"↔"Lee, Mary"), we use an unsupervised method to group all the value pairs in the same cluster (which we call *candidate replacements*) such that the value pairs in the same group can be transformed in the same way (i.e., they share a *transformation*). For example, Figure 2 shows 12 sample candidate replacements in Table 1, along with 6 groups generated by our unsupervised method. The details are given in the following sections.

Since users usually are not willing to apply a transformation blindly, we ask a human to verify each group. The human browses the value pairs in a group and marks the group as either correct (meaning the transformation is valid, with most or all value pairs representing true variant values) or incorrect (meaning the transformation is invalid and no replacement should be made). The human is not required to exhaustively check all pairs; our method is robust to small numbers of errors as verified in our experiment.

Usually there is a budget for human effort. Therefore we rank the groups by their size and ask a human to check the groups sequentially until the budget is exhausted or the Trifacta and achieved better precision (all above 99%) and recall (improved by up to 0.3) in standardizing the data with less human effort. Note that instead of verifying the transformations (i.e., groups of value pairs) in our approach, the user was asked to write code (i.e., rules) in the baseline method.

In summary, we make the following contributions:

- We propose an unsupervised method to learn string transformations for entity consolidation. We extend an existing DSL to make it more expressive and effective in producing transformations and value pair groups.

- We prove it is NP-complete to optimally group the value pairs using our criteria. We design an algorithm to greedily group the value pairs, along with optimization techniques to make it more efficient.

- We develop an incremental algorithm to efficiently generate the groups. At each invocation it guarantees to produce the next largest group for a human to verify.

- We conducted experiments on three real-world datasets. In an address dataset with 17,497 records, by having a human confirm only 100 algorithm-generated groups, we achieved a recall of 75% and a precision of 99.5% for identifying and standardizing variant value pairs.

The rest of the paper is organized as follows. Section 2 defines the problem. Section 3 presents our framework. We introduce the DSL in Section 4 and give our unsupervised string transformation learning algorithm, along with optimization techniques, in Section 5. The incremental algorithm is presented in Section 6. Section 7 discusses some implementation details. We report experimental results, review related work, and conclude in Sections 8, 9, and 10.

## 2  PROBLEM DEFINITION

Entity consolidation assumes a collection of clusters of duplicate records and produces a "golden record" for each cluster that contains the canonical value for each attribute. In this paper, we focus on the variant value standardization problem in entity consolidation, which identifies and transforms the variant values to the same format, as formalized below.

DEFINITION 1. *Given a collection* $\mathbf{C}$ *of clusters where each cluster* $C \in \mathbf{C}$ *contains a set of duplicate records, the data standardization problem in entity consolidation is to update the clusters such that all the variant values (logically the same values in different formats) in the same cluster become identical.*

As shown in Figure 1, a solution to the data standardization problem will ideally take Table 1 as input and output Table 2. In this paper, we focus on the popular case of string values. Different tactics are needed for numerical values.

---

**Algorithm 1:** GOLDENRECORDCREATION(C)

**Input**: C : a set of clusters with $m$ columns
$\quad\quad\quad \mathbf{C}^1, \mathbf{C}^2, \cdots, \mathbf{C}^m$;
**Output**: GR: a set of golden records, one for each cluster;

1 **begin**
2    **foreach** *column* $\mathbf{C}^i$ *in* $\mathbf{C}$ **do**
3       $\Phi$ = GeneratingCandidateReplacements($\mathbf{C}^i$);
4       $\Sigma_{trans}$ = UnsupervisedGrouping($\Phi$);
5       **while** *the budget is not exhausted* **do**
6          pick the largest group $\Sigma$ in $\Sigma_{trans}$;
7          **if** *a human confirms $\Sigma$ is correct* **then**
8             apply the replacements in $\Sigma$ to update $\mathbf{C}^i$;
9          remove $\Sigma$ from $\Sigma_{trans}$ and update $\Sigma_{trans}$;
10    GR = TruthDiscovery(C);
11    **return** GR
12 **end**

---

## 3  THE FRAMEWORK

Our golden record construction framework is given in Algorithm 1. It takes a set of clusters $\mathbf{C}$ as input and outputs a golden record for each cluster. At each iteration it processes one column/attribute $\mathbf{C}^i$ in $\mathbf{C}$ by the following steps.

**Step 1: Generating Candidate Replacements.** A *replacement* is an expression of the form $\texttt{lhs} \rightarrow \texttt{rhs}$ where $\texttt{lhs}$ and $\texttt{rhs}$ are two different strings. A replacement states that the two strings $\texttt{lhs}$ and $\texttt{rhs}$ are matched (e.g., "Mary Lee"→"Lee, Mary"), and thus one could be replaced by the other at certain places[1] in $\mathbf{C}^i$. Since the values within the same cluster in $\mathbf{C}^i$ are potentially duplicates, one way to get candidate replacements is to simply enumerate every pair of non-identical values $v_j$ and $v_k$ within the same cluster in $\mathbf{C}^i$ and use them to form two candidates replacements $v_j \rightarrow v_k$ and replacement $v_k \rightarrow v_j$. For example, consider the attribute Name in Table 1. We will generate 12 candidate replacements from the two clusters. 6 of them are shown in the top-left of Figure 2. Additional fine-grained token-level candidate replacement generation is given in Appendix A. By the end of this step, we have a set $\Phi$ of candidate replacements (Line 3).

**Step 2: Generating Groups of Replacements.** In this step, we partition the candidate replacements in $\Phi$ into groups such that the candidate replacements in the same group share a common transformation (which describes how one string transformed to another). We introduce a language to formally express transformations in Section 4.1. Note that each replacement group corresponds to a transformation; thus this step is essentially conducting unsupervised string transformation learning. For example, Figure 2 shows 6 groups generated from the 12 candidate replacements. Group 1 shares the same transformation of transposing the first and last name; while group 2 objects take the initial of the first name and concatenate it with the last name.

---

[1]e.g., not all "St"s are "Street" in addresses; they can also be "Saint".

**Table 4: Sample groups generated by our unsupervised method from a real-world book-author-list dataset**

| Group A | Group B | Group C |
|---|---|---|
| "fox, dan box, jon"→"dan fox, jon box" | "bobby"→"bob" | "knuth, donald e."→"donald e. knuth" |
| "egan, mark mather, tim"→"mark egan, tim mather" | "jeffrey"→"jeff" | "hutton, david v."→"david v. hutton" |
| "irvine, kip gaddis, tony"→"kip irvine, tony gaddis" | "matthew"→"matt" | "nilsson, nils j."→"nils j. nilsson" |
| "parr, mike bell, douglas"→"mike parr, douglas bell" | "steven"→"steve" | "thomas w. miller"→"miller, thomas w." |
| "gray, jim reuter, andreas"→"jim gray, andreas reuter" | "kenneth"→"ken" | "judith s. bowman"→"bowman, judith s." |

| Group D | Group E |
|---|---|
| "levy, margipowell, philip"→"margi levy, philip powell" | "carroll, john (edt)"→"john carroll" |
| "bohl, marilynrynn, maria"→"marilyn bohl, maria rynn" | "williams, jim (edt)"→"jim williams" |
| "arthorne, johnlaffra, chris"→"john arthorne, chris laffra" | "brown, keith (author)"→"keith brown" |
| "langer, angelikakreft, klaus"→"angelika langer, klaus kreft" | "wagner, bill (author)"→"bill wagner" |
| "kroll, permacisaac, bruce"→"per kroll, bruce macisaac" | "lieberman, henry (editor)"→"henry lieberman" |

As another motivating example, we demonstrate 5 example groups that our method generated from a real-world book-author-list dataset[2] in Table 4. Note that for each group, we only show 5 candidate replacements out of hundreds due to space limitation. Though it is not shown in the table, all the candidate replacements in each group share a transformation. For example, for group A, the author lists on the left-hand side use whitespace as separators for names while those on the right-hand side use commas. Also the orders of the first and last names are reversed.

By the end of this step, we have a set $\Sigma_{trans}$ of groups, which is a partition of $\Phi$ (Line 4).

**Step 3: Applying Approved Replacement Groups.** The groups in $\Sigma_{trans}$ are ranked by their size in descending order. We sequentially present each replacement group to a human expert for verification. The expert either rejects or approves a replacement group. If it is approved, the expert needs to further specify the replacement direction, i.e., either replacing `lhs` with `rhs` or the other way around. The verification stops once the budget is exhausted or the expert is satisfied with the results.

The reason for confirming the groups in decreasing size order is twofold. First, the more replacements there are in a group, the more places we can apply them to update the clusters once the group is approved by the human, i.e., the larger groups are more 'profitable' once they are approved. Second, the larger groups are more likely to be approved, as variant values often share common transformations that appear repeatedly across different clusters (e.g., transposing the first and last name). On the other hand, the transformations of value pairs in smaller groups are more peculiar and uncommon. Thus the value pairs in smaller groups are less likely to be variant values and get approved. Section 7.1 gives the details. By the end of this step, $C^i$ is updated (Lines 5-9).

**Running Truth Discovery.** Finally, after we process all the columns in C by the above steps, a truth discovery algorithm is applied on the updated clusters C to resolve any potential conflicts. In the end, we have the golden records (Line 10).

## 4 GENERATING GROUPS

We introduce the DSL in Section 4.1 and discuss how to group the candidate replacements in $\Phi$ in Section 4.2.

### 4.1 Transformation Programs

**Transformation Programs.** A *transformation program* (or program for short) captures how one string is transformed to another. We adopt the domain specific language (DSL) designed by Gulwani [22, 23] to express the programs, which we formally summarize in Appendix B. Here we give a high level description. In a nutshell, a transformation program takes a string **s** as input and outputs another string **t**. The DSL defines *position functions* and *string functions*, which all apply to the input string **s**.

A position function returns an integer that indicates a position in **s** based on a collection of pre-defined regular expressions (regexes). For example, Figure 3 on the left shows some example position functions and pre-defined regexes. Let the input string **s** be "Lee, Mary", as shown in Figure 4, we have $\mathbf{s}.P_A = 1$ as the $1^{st}$ match of the capital regex $\mathsf{T}_C$ in **s** is "L" and the beginning position of "L" is 1.

A string function returns either a substring of **s** or a constant string. The returned substring is located by position functions. A program is defined as a sequence of string functions and its output **t** is the concatenation of the outputs of these string functions. Figures 3 and 4 show an example. For

| | | | |
|---|---|---|---|
| capital regex: $T_C$= [A-Z]+ | $P_A$ : beginning of the $1^{st}$ match of $T_C$ | $f_1$ : Substring ($P_A$, $P_B$) | |
| lowercase regex: $T_l$= [a-z]+ | $P_B$ : ending of the $1^{st}$ match of $T_l$ | $f_2$ : Substring ($P_C$, $P_D$) | $\rho := f_2 \oplus f_3 \oplus f_1$ |
| whitespace regex: $T_b$ = \s+ | $P_C$ : ending of the $1^{st}$ match of $T_b$ | $f_3$ : Constant (". ") | |
| digital regex: $T_d$ = [0-9]+ | $P_D$ : ending of the last (-$1^{st}$) match of $T_C$ | | |
| example pre-defined **regexes** | example **position functions** | example **string functions** | an example **program** |

**Figure 3: An example program $\rho := f_2 \oplus f_3 \oplus f_1$**

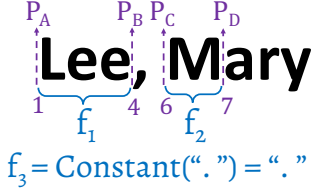

$f_3$ = Constant(". ") = ". "

$\rho$("Lee, Mary") = $f_2 \oplus f_3 \oplus f_1$ = "M. Lee"

**Figure 4: Evaluating the example functions and program $\rho$ in Figure 3 on an input string "Lee, Mary"**

$\mathbf{s}$ = "Lee, Mary", we have $\mathbf{s}.f_1$ = "Lee" and $\mathbf{s}.f_2$ = "M". The program $\rho := f_2 \oplus f_3 \oplus f_1$ will produce $\mathbf{t} = \rho(\mathbf{s})$ = "M. Lee".

**Transformation Graph.** We say a program $\rho$ is consistent with a replacement $\mathbf{s} \rightarrow \mathbf{t}$ (or $\rho$ can express the replacement) iff $\rho(\mathbf{s})$ produces $\mathbf{t}$. Due to the many possible combinations of string functions, there are an exponential number of *consistent programs* for a given replacement $\mathbf{s} \rightarrow \mathbf{t}$. Fortunately, all the consistent programs of a replacement can be encoded in a directed acyclic graph (DAG) in polynomial time and space [22, 32]. Intuitively, each node in the graph corresponds to a position in $\mathbf{t}$, each edge in the graph corresponds to a substring of $\mathbf{t}$ and the labels on the edge are string functions that return this substring when being applied to $\mathbf{s}$.

*Example 4.1.* Figure 5 shows the transformation graph for "Lee, Mary"→"M. Lee". Some notations are borrowed from Figure 3. The string functions (edge labels) associated with each edge are also shown in the figure. For simplicity, we only show 5 out of all the 21 edges and ignore some edge labels in the figure. The edge $e_{4,7}$ corresponds to the substring "Lee". One of its label is $f_1$, as it returns "Lee" when being applied to $\mathbf{s}$ = "Lee, Mary" as discussed before. Substring($P_A$, $P_E$) is also a label of $e_{4,7}$, as $\mathbf{s}.P_A$ = 1, $\mathbf{s}.P_E$ = 4, and $\mathbf{s}$.Substring(1, 4) = "Lee".

Formally, the transformation graph is defined as below.

**Definition 2 (Transformation Graph).** *Given a replacement* $\mathbf{s} \rightarrow \mathbf{t}$, *its transformation graph is a directed acyclic graph* $G(\mathbf{N}, \mathbf{E})$ *with a set* $\mathbf{N}$ *of nodes and a set* $\mathbf{E}$ *of edges. There are* $|\mathbf{t}| + 1$ *nodes, i.e.,* $\mathbf{N} = \{n_1, \ldots, n_{|\mathbf{t}|+1}\}$. *There is a directed edge* $e_{i,j} \in \mathbf{E}$ *from* $n_i$ *to* $n_j$ *for any* $1 \leq i < j \leq |\mathbf{t}| + 1$. *Moreover, each edge* $e_{i,j}$ *is labeled with a set of string functions that returns* $\mathbf{t}[i, j-1]$ *when being applied to* $\mathbf{s}$.

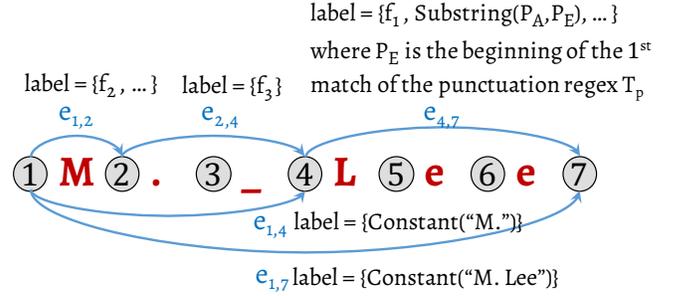

**Figure 5: The graph for "Lee, Mary"→"M. Lee"**

The transformation graph construction algorithm is given in Appendix C. It runs in $O(|\mathbf{s}|^2|\mathbf{t}|^2)$ time and there are $O(|\mathbf{t}|^2)$ edges in the graph. As a replacement has only one transformation graph, we refer to a replacement and its transformation graph (or graph for short) interchangeably.

**Transformation Path.** Given a replacement $\mathbf{s} \rightarrow \mathbf{t}$, a *transformation path* is a path in its graph from the *first node* $n_1$ to the *last node* $n_{|\mathbf{t}|+1}$, where each edge has only one label (i.e., a string function). Note that a transformation path uniquely refers to a consistent program. We use them interchangeably. For instance, two transformation paths in Figure 5 are:

$\rho_1 := ① \xrightarrow{f_2} ② \xrightarrow{f_3} ④ \xrightarrow{f_1} ⑦$ and $\rho_2 := ① \xrightarrow{\text{Constant("M.Lee")}} ⑦$

Given a transformation $\phi$, all the transformation paths in its graph are consistent with $\phi$ and all the programs consistent with $\phi$ are within its graph as stated below.

**Theorem 4.2.** *Given a transformation, there is a bijection between its consistent programs and the transformation paths in its transformation graph [22].*

Based on the above theorem, we can use the set $\mathbf{G}$ of graphs corresponding to the replacements in $\Phi$ to group the candidate replacements in $\Phi$. Next we discuss the details.

## 4.2 The Optimal Partition Problem

As discussed before, given a collection of replacements, we aim to group them such that the replacements within the same group share a consistent program (i.e., their graphs share a transformation path) while the number of groups is minimum. We formalize the problem as below.

**Definition 3 (Optimal Partition).** *Given a set* $\Phi$ *of replacements, the* optimal partition *problem is to partition* $\Phi$ *into disjoint groups* $\Phi_1, \ldots, \Phi_n$ *such that (i) the replacements* *in each group $\Phi_i$ share at least one consistent program; and (ii) the number of groups $n$ is minimum.*

Unfortunately, we can prove the optimal partition problem is NP-complete by a reduction from the set cover problem (proof sketch: each transformation path corresponds to the set of graphs in $\Phi$ containing this path; the optimal partition problem is to find the minimum number of transformation paths such that the set of all graphs is covered). As it is prohibitively expensive to find the optimal partition, we employ a standard greedy strategy. For each replacement $\varphi \in \Phi$, we denote the transformation path in its graph that is shared by the largest number of the graphs in $\Phi$ as its *pivot path*. Then the replacements with the same pivot path are grouped together[3]. We discuss how to find the pivot path for a given replacement in $\Phi$ in the next section. In this way, we can partition the replacements in $\Phi$ into disjoint groups.

## 5 SEARCHING FOR THE PIVOT PATH

We use $\mathbf{G}$ to denote the set of graphs corresponding to the candidate replacements in $\Phi$. Section 5.1 gives the pivot path search algorithm and Section 5.2 presents two optimizations.

### 5.1 Pivot Path Search Algorithm

A naive method enumerates every transformation path $\rho$ in a graph $G$ and counts the number of graphs in $\mathbf{G}$ containing $\rho$. Then the pivot path is the transformation path contained by the largest number of graphs. For this purpose, given a path $\rho := f_1 \oplus f_2 \oplus \cdots \oplus f_m$, we first show how to get the list of graphs in $\mathbf{G}$ containing $\rho$.

We observe that if a graph $G$ contains $\rho$, every string function $f_1, f_2, \cdots, f_m$ of $\rho$ must appear in $G$ as a label. Thus we can build an inverted index $I$ with string functions as keys. The inverted list $I[f]$ consists of all the graphs $G \in \mathbf{G}$ that have $f$ (i.e., $f$ is an edge label in $G$). Then given a path $\rho := f_1 \oplus \cdots \oplus f_m$, we can find the list of graphs in $\mathbf{G}$ that contain $\rho$ by taking the intersection $I[f_1] \cap \cdots \cap I[f_m]$.

However, since $\rho$ is a path, the edges corresponding to the string functions $f_1, f_2, \cdots, f_m$ are required to be adjacent in the graphs. To enable this, we also add the edge information to the entries of the inverted lists. In particular, the inverted list $I[f]$ consists of all triples $\langle G, i, j \rangle$ such that the edge $e_{ij}$ from $n_i$ to $n_j$ in $G$ has the label $f$. Then, when intersecting $I[f_1]$ with $I[f_2]$, only if an entry $\langle G, i_1, j_1 \rangle$ from $I[f_1]$ and another entry $\langle G, i_2, j_2 \rangle$ from $I[f_2]$ satisfy $j_1 = i_2$ (i.e., their edges $e_{i_1, j_1}$ and $e_{i_2, j_2}$ are adjacent), they produce a new entry $\langle G, i_1, j_2 \rangle$ in the result list. By doing so, one can verify that $I[f_1] \cap \cdots \cap I[f_m]$ is exactly the list of graphs in $\mathbf{G}$ containing $\rho$. Hereinafter, whenever we intersect two lists, we intersect them in the way we described above.

---

[3]Note that two paths are considered to be the same if each pair of string functions in the two sequences are the same.

---

**Algorithm 2:** UNSUPERVISEDGROUPING($\Phi$)

**Input**: $\Phi$: a collection of candidate replacements.
**Output**: $\Sigma$: groups of replacements with the same transformation, where $\Sigma[\rho]$ contains all the replacements in $\Phi$ with $\rho$ as the pivot path.

1 **begin**
2     Build graphs $\mathbf{G}$ for all replacements in $\Phi$;
3     Build inverted index $I$ for all edge labels in $\mathbf{G}$;
4     **foreach** *graph* $G \in \mathbf{G}$ **do**
5         $\rho = \rho_{max} = \ell_{max} = \phi$;
6         SEARCHPIVOT($G$, $\mathbf{G}$, $n_1$, $\rho_{max}$, $\ell_{max}$);
7         add $G$ to the group $\Sigma[\rho_{max}]$;
8     **return** $\Sigma$;
9 **end**

---

**Algorithm 3:** SEARCHPIVOT($\rho$, $\ell$, $n_i$, $\rho_{max}$, $\ell_{max}$)

**Input**: $G$: a transformation graph;
$\rho$: a path in $G$ starting from $n_1$;
$\ell$: the list of graphs in $\mathbf{G}$ containing $\rho$;
$n_i$: the node at the end of $\rho$;
$\rho_{max}$: the best path in $G$ found so far;
$\ell_{max}$: the list of graphs in $\mathbf{G}$ containing $\rho_{max}$.

1 **begin**
2     **if** $n_i$ *is the last node in $G$* **then**
3         **if** $|\ell| > |\ell_{max}|$ **then**
4             $\rho_{max} = \rho$;
5             $\ell_{max} = \ell$;
6     **else**
7         **foreach** *edge $e_{i,j}$ from $n_i$ to $n_j$ in $G$* **do**
8             **foreach** *string function label $f$ on $e_{i,j}$* **do**
9                 $\rho' = \rho \oplus f$;
10                 $\ell' = \ell \cap I[f]$;
11                 SEARCHPIVOT($G$, $\rho'$, $\ell'$, $n_j$, $\rho_{max}$, $\ell_{max}$);
12 **end**

---

**Algorithm 4:** EARLYTERMINATION

1 **begin**
    // add after Line 2 of Algorithm 2
2     **foreach** *graph $G \in \mathbf{G}$* **do** set $G_{lo}$ as 1;
    // add after Line 2 of Algorithm 3
3     **foreach** *graph $G' \in \ell$* **do**
4         **if** $G'_{lo} < |\ell|$ **then** $G'_{lo} = |\ell|$;
    // add before Line 11 of Algorithm 3
5     **if** $|\ell'| > |\ell_{max}|$ *and* $|\ell'| \geq G_{lo}$ **then**
6 **end**

**Table 5: An Example of SearchPivot**

| | $\rho$ | $\ell$ | $n_i$ | $\rho_{max}$ | $\ell_{max}$ | $f$ |
|---|---|---|---|---|---|---|
| 1 | $\phi$ | $\{\langle G_1, 1, 1\rangle, \langle G_2, 1, 1\rangle, \langle G_3, 1, 1\rangle\}$ | $n_1$ | $\phi$ | $\phi$ | Constant("M. Lee") on $e_{1,7}$ |
| 2 | Constant("M. Lee") | $\{\langle G_1, 1, 7\rangle\}$ | $n_7$ | Constant("M. Lee") | $\{\langle G_1, 1, 7\rangle\}$ | — |
| 3 | $\phi$ | $\{\langle G_1, 1, 1\rangle, \langle G_2, 1, 1\rangle, \langle G_3, 1, 1\rangle\}$ | $n_1$ | Constant("M. Lee") | $\{\langle G_1, 1, 7\rangle\}$ | $f_2$ on $e_{1,2}$ |
| 4 | $f_2$ | $\{\langle G_1, 1, 2\rangle, \langle G_2, 1, 2\rangle\}$ | $n_2$ | Constant("M. Lee") | $\{\langle G_1, 1, 7\rangle\}$ | $f_3$ on $e_{2,4}$ |
| 5 | $f_2 \oplus f_3$ | $\{\langle G_1, 1, 4\rangle, \langle G_2, 1, 4\rangle\}$ | $n_4$ | Constant("M. Lee") | $\{\langle G_1, 1, 7\rangle\}$ | $f_1$ on $e_{4,7}$ |
| 6 | $f_2 \oplus f_3 \oplus f_1$ | $\{\langle G_1, 1, 7\rangle, \langle G_2, 1, 9\rangle\}$ | $n_7$ | $f_2 \oplus f_3 \oplus f_1$ | $\langle G_1, 1, 7\rangle, \langle G_2, 1, 9\rangle$ | — |

*Example 5.1.* Consider $\Phi = \{\varphi_1, \varphi_2, \varphi_3\}$ where $\varphi_1$ ="Lee, Mary"→"M. Lee", $\varphi_2$="Smith, James"→"J. Smith", and $\varphi_3$="Lee, Mary"→"Mary Lee". Let $G_1$, $G_2$, and $G_3$ be the transformation graphs of $\varphi_1, \varphi_2$, and $\varphi_3$, respectively. Using the string functions $f_1, f_2$, and $f_3$ in Figures 3-5, we have $I[f_1] = (\langle G_1, 4, 7\rangle, \langle G_2, 4, 9\rangle, \langle G_3, 6, 9\rangle)$, $I[f_2] = (\langle G_1, 1, 2\rangle, \langle G_2, 1, 2\rangle, \langle G_3, 1, 2\rangle)$, and $I[f_3] = (\langle G_1, 2, 4\rangle, \langle G_2, 2, 4\rangle)$. The path $f_2 \oplus f_3 \oplus f_1$ is contained by $\varphi_1$ and $\varphi_2$ as $I[f_2] \cap I[f_3] \cap I[f_1] = (\langle G_1, 1, 7\rangle, \langle G_2, 1, 9\rangle)$.

**The Search Algorithm.** As there are an exponential number of transformation paths in a graph, it is prohibitively expensive for the naive method to enumerate all of them. To alleviate this problem, we give a recursive algorithm to find the pivot path in a graph $G$. At a high level, the algorithm maintains a path $\rho$ in $\mathbf{G}$ starting from the first node $n_1$ and the list $\ell$ of all graphs in $\mathbf{G}$ containing $\rho$. At each invocation, the algorithm will try to append a label (string function) $f$ on the outgoing edges of the last node $n_i$ in $\rho$ to the end of $\rho$ and update $\ell$ to the list of graphs containing the new path. After this, if $\rho$ does not reach the last node in $G$, the algorithm will be invoked again to further extend $\rho$. Otherwise $\rho$ reaches the last node and must be a transformation path; and $\rho$ is the pivot path if $\ell$ has the largest number of the graphs in $\mathbf{G}$.

Algorithm 3 gives the pseudo-code of the search algorithm. At each invocation, it takes six parameters: a graph $G$, a path $\rho$ in $G$ starting from the first node $n_1$, the list $\ell$ of graphs in $\mathbf{G}$ containing $\rho$, the node $n_i$ at the end of $\rho$, the best path $\rho_{max}$ (i.e., contained by the largest number of graphs in $\mathbf{G}$) in $G$ found so far, and the list $\ell_{max}$ of graphs in $\mathbf{G}$ containing $\rho_{max}$. First, it checks whether the maintained path $\rho$ is a transformation path (Line 2). If $\rho$ is a transformation path and there are more graphs in $\mathbf{G}$ containing $\rho$ than $\rho_{max}$, the algorithm updates the best path $\rho_{max}$ found so far as $\rho$ and the list $\ell_{max}$ as $\ell$ (Lines 3 to 5). If $\rho$ is not a transformation path, it tries to extend $\rho$ with one more edge label (string function). Specifically, for each outgoing edge $e_{i,j}$ of the node $n_i$, which must be adjacent to $\rho$, and each string function label $f$ on $e_{i,j}$, it appends $f$ to the end of $\rho$ to get a new path $\rho'$ and intersects $\ell$ with the inverted list $I[f]$ to get the list $\ell'$ of graphs containing the new path $\rho'$ (Lines 7 to 10). Then the algorithm recursively invokes itself to examine the new path $\rho'$ (Line 11). When the recursive algorithm terminates, $\rho_{max}$ must be the pivot path of $G$. Initially, $\rho$, $\rho_{max}$, and $\ell_{max}$ are all empty while $\ell$ contains all the graphs in $\mathbf{G}$, as an empty path can be contained by any graph (Line 5 of Algorithm 2).

*Example 5.2.* Consider the graphs $\mathbf{G}$ in Example 5.1. We invoke SearchPivot to search the pivot path of $G_1$. As shown in the first row of Table 5, initially $\rho$, $\rho_{max}$, and $\ell_{max}$ are all $\phi$ and $\ell$ has all the graphs in $\mathbf{G}$. Next we go through every label on the edges starting from $n_1$.

For example, consider the label Constant("M. Lee") on $e_{1,7}$ as shown in Figure 5. We update the maintained path $\rho$ and list $\ell$, as shown in row 2 of Table 5, and invoke Search­Pivot again with $n_i$ as the endpoint of $e_{1,7}$, i.e., $n_7$. Since $n_7$ is the last node in the graph, $\rho$ is a transformation path and we assign $\rho$ and $\ell$ to $\rho_{max}$ and $\ell_{max}$ respectively.

Next, we explore another edge starting from $n_1$. The maintained path $\rho$ and list $\ell$ are restored as shown in row 3 of Table 5. Consider the label $f_2$ on $e_{1,2}$. We update $\rho$ and $\ell$ to row 4 of Table 5 and invoke SearchPivot again with $n_i$ as the endpoint of $e_{1,2}$, i.e., $n_2$. As $n_2$ is not the last node in $G_1$, we further go through the labels on the edges starting from $n_2$. Eventually, as shown in row 6 of Table 5 , $\rho$ is extended to a transformation path $f_2 \oplus f_3 \oplus f_1$. Since the list $\ell$ has more graphs than $\ell_{max}$, we update $\rho_{max}$ and $\ell_{max}$ by $\rho$ and $\ell$ respectively. The algorithm continues to explore the graph but could not find any better transformation path. Finally it returns the pivot path $\rho_{max} = f_2 \oplus f_3 \oplus f_1$.

Algorithm 2 gives the pseudo-code of our unsupervised string transformation learning algorithm Unsupervised­Grouping. Each generated group $\Sigma[\rho]$ corresponds to a string transformation program $\rho$ written in our DSL.

## 5.2 Improving Pivot Path Search

In this section, we introduce two optimizations to improve the efficiency of the pivot path search algorithm. Intuitively, intersecting two inverted lists cannot result in a longer list. Thus if the length of $\ell$ is no longer than that of $\ell_{max}$, we can skip recursively invoking the algorithm to extend $\rho$ to a transformation path as it cannot result in any transformation path contained by more graphs in $\mathbf{G}$ than $\rho_{max}$ (i.e., cannot result in the pivot path). Next we discuss the details.

**Local Threshold-based Early Termination.** The length of the maintained list $\ell$ decreases monotonically as the maintained path $\rho$ getting longer. This is because each time a label $f$ is appended to $\rho$, $\ell$ is updated to $\ell \cap I[f]$ and gets shorter.

As we only need the pivot path – the one that is shared by the largest number of the graphs in $\mathbf{G}$, we can use $|\ell_{max}|$ as a (local) threshold: only if $|\ell| > |\ell_{max}|$, the algorithm is recursively invoked. To enable this, we add an if condition "$|\ell'| > |\ell_{max}|$" before Line 11 of Algorithm 3.

**Global Threshold-based Early Termination.** Once the maintained path $\rho$ becomes a transformation path (Line 2 in Algorithm 3), we know all the graphs in $\ell$ must contain $\rho$. Thus, for each graph in $\ell$, any path no better than $\rho$ must not be its pivot path. Specifically, we can use $|\ell|$ as a (global) threshold for those graphs in $\ell$. Then when searching for the pivot paths of the graphs in $\ell$, we can use this global threshold for early termination in the same way as the local threshold. To enable this in Algorithm 3, we can associate each graph $G$ with a global threshold $G_{lo}$. Then whenever a transformation path is found (Line 2 in Algorithm 3), the corresponding global threshold $G'_{lo}$ of the graph $G'$ in $\ell'$ will be updated to $|\ell'|$ if $|\ell'|$ is larger. Algorithm 4 shows how to enable the two early optimizations in SearchPivot.

*Example 5.3.* Continue with Example 5.2. At the $6^{th}$ row of Table 5, we have the local threshold of $G_1$ as $|\ell_{max}| = 2$. Then, as none of the edge labels of $G_1$ has an inverted list length longer than 2, we will not invoke SearchPivot any more and have the pivot path $\rho_{max} := f_2 \oplus f_3 \oplus f_1$. Moreover, as $\rho$ is a transformation path, we set the global threshold of $G_2 \in \ell$ as $|\ell| = 2$. Then, when we search for the pivot path of $G_2$, we can skip all edge labels with inverted list length shorter than 2, including the `Constant("J. Smith")` on $e_{1,9}$.

## 6 INCREMENTAL GROUPING METHOD

We observe that the approach UnsupervisedGrouping in Section 5.1 partitions all the replacements upfront. This will incur a huge upfront cost, i.e., the users need to wait a long time before any group is generated. Moreover, due to the limited budget, many small groups will not be presented to the user for verification and it is unnecessary to generate them. To alleviate this problem, we propose an incremental algorithm (i.e., top-k grouping) in this section. It produces the largest group at each invocation. Next we give the details.

### 6.1 Largest Group Generation

We first give the intuition of largest group generation. We denote the pivot path that is shared by the largest number of graphs in $\mathbf{G}$ as the *best pivot path* $\rho_{best}$. Then the list $\ell_{best}$ of graphs containing $\rho_{best}$ must be the largest group. This is because no other path can be shared by more graphs than $\rho_{best}$. Next we show how to calculate $\rho_{best}$ and $\ell_{best}$.

Intuitively, for each graph $G$ in $\mathbf{G}$, we associate it with an upper bound and a lower bound of the number of graphs in $\mathbf{G}$ containing its pivot path. Let $\tau$ be the largest lower bound among all the graphs in $\mathbf{G}$. We visit each graph in

$\mathbf{G}$ and invoke SearchPivot to find its pivot path. In this process, the lower and upper bounds of the graphs in $\mathbf{G}$ will become tighter and the largest lower bound $\tau$ will be updated accordingly. We stop once $\tau$ is no smaller than any upper bound of the unvisited graphs. Then the pivot path we found so far that is shared by the largest number of graphs must be the best pivot path $\rho_{best}$. This is because for the unvisited graphs, their pivot paths must be shared by no more than $\tau$ graphs, while one of the pivot paths in the visited graphs must be shared by no less than $\tau$ graphs.

Formally, as discussed in Section 5.2, for a graph $G$, we update its global threshold $G_{lo}$ only if a transformation path in $G$ is found by SearchPivot. Thus we can use the global threshold $G_{lo}$ as the lower bound of $G$. We discuss how to initialize the upper bound $G_{up}$ in Section 6.2. Then we sort all the graphs in $\mathbf{G}$ by their upper bounds in descending order and visit them sequentially. As we only need the best pivot path $\rho_{best}$, it is unnecessary to find the pivot path of a graph $G$ if it is shared by no more than $\tau$ graphs in $\mathbf{G}$. For this purpose, when visiting a graph $G$ and searching for its pivot path, we use $\tau$ as a local threshold (recall Section 5.2) in SearchPivot. Then SearchPivot either finds its pivot path shared by more than $\tau$ graphs or concludes its pivot path cannot be shared by more than $\tau$ graphs. In the latter case, we can assign $\tau$ as a tighter upper bound for $G$. In the former case, since SearchPivot already finds the pivot path $\rho_{max}$ of $G$, we can update its lower and upper bound to the number of graphs sharing $\rho_{max}$. Moreover, since $\rho_{max}$ is shared by more than $\tau$ graphs, the largest lower bound $\tau$ should also be updated. We stop whenever $\tau$ is no smaller than the upper bound of the currently visiting graph as the graphs are ordered by their upper bounds. Then all the graphs with lower bounds equal to $\tau$ form the largest group.

*Example 6.1.* Continue with Example 5.1. Initially the lower bounds of $G_1$, $G_2$, and $G_3$ are all 1 and the upper bounds are 2, 2, and 1 respectively (we will discuss this details in the next section). The largest lower bound $\tau = 1$. Then we invoke SearchPivot to find the pivot path of $G_1$ as discussed in Example 5.2. We find the pivot path $\rho_{max}$ of $G_1$ is shared by 2 graphs $G_1$ and $G_2$. Thus we update the lower bound of $G_1$ to 2 and the largest lower bound $\tau$ to 2. Next we visit the second graph $G_2$. Since its upper bound is 2, which is no larger than $\tau = 2$, we can stop and $\rho_{max}$ must be the best pivot path and the largest group consists of $G_1$ and $G_2$.

### 6.2 Initializing the Upper Bounds

Next we discuss how to initialize an upper bound for a graph. We observe that the pivot path is a sequence of edge labels and the number of graphs sharing the pivot path is the intersection size of the inverted lists of these edge labels. Thus, for any graph, we can use the length of the longest inverted list among all its edge labels as an upper bound as the intersection cannot result in longer lists.

Clearly, we desire the upper bound to be as tight as possible to reach the stop condition earlier. To achieve a tighter upper bound, we have the following observation. The pivot path must cover the entire output string $t$, i.e., it goes from the first node $n_1$ to the last $n_{|t|+1}$. Thus, for any node $n_k$, one of the edges $e_{i,j}$ where $i \leq k < j$ (i.e., $e_{i,j}$ "covers" $n_k$) must appear in the pivot path. Based on this observation, we can deduce an upper bound $ub[k]$ from any position $n_k$, which is the length of the longest inverted list among all the labels of the edge $e_{i,j}$ $i \leq k < j$, i.e., $e_{i,j}$ covers $n_k$.

LEMMA 6.2. *Consider a graph $G \in \mathbf{G}$, for any $1 \leq i \leq |t|$, $ub[i]$ is an upper bound of the number of graphs in $\mathbf{G}$ sharing the pivot path of $G$.*

Since every value in $ub$ is an upper bound, we use the tightest one (i.e., the smallest value in $ub$) to initialize $G_{up}$.

*Example 6.3.* Continue with Example 5.1. For $G_1$, we have $ub[5] = 3$ as the label $f_1$ on $e_{4,7}$ has an inverted list $I[f_1]$ of length 3 as shown in Example 5.1. $ub[2] = 2$ as none of the labels of $G_3$ can produce the character '.'. Finally the upper bound of $G_1$ is initialized as $ub[2] = 2$.

## 6.3 The Incremental Algorithm

Algorithm 5 gives the pseudo-code of our incremental algorithm. Instead of invoking UNSUPERVISEDGROUPING in our framework as shown in Algorithm 1, we first invoke Algorithm 6 to preprocess the candidate replacements (Line 1). Then, while the budget is not exhausted, we invoke Algorithm 7 to produce the next largest group $\Sigma$ for a human to verify (Line 2) and update the graphs as necessary (Line 3).

Algorithm 6 takes the set $\Phi$ of candidate replacements as input. It creates the graphs $\mathbf{G}$ for $\Phi$ (Line 2), builds the inverted index $I$ (Line 3), and initializes the lower bounds (Line 5) and upper bounds (Lines 6-11) for the graphs in $\mathbf{G}$.

Each invocation of Algorithm 7 produces the next largest group. It first initializes the largest lower bound $\tau$ (Line 2). Then it sorts all the graphs in $\mathbf{G}$ by their upper bounds in descending order and visit them sequentially (Lines 3-4). It uses two variable $\rho_{best}$ and $\ell_{best}$ to keep the best pivot path found so far and the list of graphs in $\mathbf{G}$ containing $\rho_{best}$. When visiting a graph $G$, it first checks whether its upper bound is larger than $\tau$. If so, we can stop and return $\ell_{best}$ as $\rho_{best}$ must be the best pivot path (Line 5). Otherwise, it invokes SEARCHPIVOT to check if the pivot path of $G$ is contained by more than $\tau$ graphs. For this purpose, it initializes $\ell_{max}$ with $\tau$ random graph such that the local threshold $|\ell_{max}| = \tau$ (Lines 6-8). In this way, only if the maintained path $\rho$ is shared by more than $\tau$ graphs (i.e., $|\ell| > \tau$), SEARCHPIVOT will be recursively invoked and $\ell_{max}$ will be updated . Then, if SEARCHPIVOT finds a pivot path $\rho_{max}$, it updates $\rho_{best}$

---

**Algorithm 5: INCREMENTALGROUPING**

  // replace Line 4 of Algorithm 1
1  $\mathbf{G}$ = Preprocessing($\Phi$);
  // replace Line 6 of Algorithm 1
2  $\Sigma$ = GenerateNextLargestGroup($\mathbf{G}$);
  // replace Line 9 of Algorithm 1
3  remove all the graphs in $\Sigma$ from $\mathbf{G}$ and update $\mathbf{G}$;

---

**Algorithm 6: PREPROCESSING($\Phi$)**

**Input**: $\Phi$: a collection of candidate replacement.
**Output**: $\mathbf{G}$: the set of graphs corresponding to $\Phi$.
1  **begin**
2     Construct graphs $\mathbf{G}$ for all replacement in $\Phi$;
3     Build inverted index $I$ for all edge labels in $\mathbf{G}$;
4     **foreach** *graph $G \in \mathbf{G}$* **do**
5        set $G_{lo}$ as 1;
6        **foreach** *edge $e_{i,j}$ in $G$* **do**
7           **foreach** *string function label $f$ on $e_{i,j}$* **do**
8              **foreach** $i \leq k < j$ **do**
9                 **if** $ub[k] < |I[f]|$ **then**
10                   $ub[k] = |I[f]|$;
11     set $G_{up}$ as the smallest value in $ub$;
12     **return** $\mathbf{G}$;
13  **end**

---

**Algorithm 7: GENERATENEXTLARGESTGROUP($\mathbf{G}$)**

**Input**: $\mathbf{G}$: a set of transformation graphs.
**Output**: $\ell_{best}$: the list of graphs in $\mathbf{G}$ containing the best path $\rho_{best}$ that shared by the largest number of graphs in $\mathbf{G}$.
1  **begin**
2     let $\tau$ be the largest lower bound in $\mathbf{G}$;
3     sort the graphs in $\mathbf{G}$ by their upper bounds descendingly;
4     **foreach** *graph $G \in \mathbf{G}$* **do**
5        **if** $\tau \geq G_{up}$ **then** Break;
6        $\rho = \rho_{max} = \phi$ ;
7        initial $\ell_{max}$ with $\tau$ random graphs s.t. $|\ell_{max}| = \tau$;
8        SEARCHPIVOT($G, \rho, \mathbf{G}, n_1, \rho_{max}, \ell_{max}$);
9        **if** $\rho_{max} \neq \phi$ **then**
10           update $G_{lo}, G_{up}$, and $\tau$ all as $|\ell_{max}|$;
11           $\rho_{best} = \rho_{max}$;
12           $\ell_{best} = \ell_{max}$;
13        **else**
14           $G_{up} = \tau$;
15     **return** $\ell_{best}$;
16  **end**

and $\ell_{best}$ as $\rho_{max}$ and $\ell_{max}$ respectively. The lower bound $G_{lo}$, upper bound $G_{up}$, and the largest lower bound $\tau$ are all updated to $|\ell_{max}|$ (Lines 10-12). Note that SearchPivot may update the lower bounds of the other graphs. However, in this case, $\tau$ is still the largest lower bound as none of the updated lower bounds can exceed $|\ell_{max}|$. If SearchPivot does not find the pivot path, it means the pivot path in $G$ cannot be shared by more than $\tau$ graphs. Thus we update $G_{up}$ as $\tau$ (Line 14). Similarly, in this case, none of the updated lower bound can be larger than $\tau$ and $\tau$ remains the largest lower bound. Finally, when the stop condition is satisfied, $\rho_{max}$ must be the best pivot path and $\ell_{max}$ must be the largest group and thus get returned (Line 15).

Theorem 6.4. *Let* $\Sigma_1, \Sigma_2, \cdots, \Sigma_m$ *be the ordered replacement groups generated by the algorithm* UnsupervisedGrouping, *where* $|\Sigma_1| < |\Sigma_2| < \cdots < |\Sigma_m|$. *The algorithm* GenerateNextLargestGroup *will return* $\Sigma_i$ *at its* $i^{th}$ *invocation.*

# 7 IMPLEMENTATION DETAILS

## 7.1 Applying Approved Groups

Once a replacement $\varphi$ is approved, we backtrack all the value pairs that generate $\varphi$ and make the change (i.e., replace one value with the other one). For this purpose, for each candidate replacement lhs → rhs, we build a *replacement set*, denote as $L[\text{lhs} \rightarrow \text{rhs}]$, to keep all the places where the replacement is generated from. In addition, after updating a value, the replacements generated from the value may change. For example, consider the three values $v_1 = r_1[\text{Name}]$, $v_2 = r_2[\text{Name}]$, and $v_3 = r_3[\text{Name}]$ in Table 1. They will generate 6 replacements. Suppose the replacement $v_1 \rightarrow v_2$ is approved and $v_1$ is replaced by $v_2$. Then, the replacement $v_1 \rightarrow v_3$ will become $v_2 \rightarrow v_3$. Moreover, the replacement $v_2 \rightarrow v_1$ no longer exists. Thus we also need to update the replacement sets after making changes to values. Next we discuss the details.

**Building Replacement Sets.** Let $v_j^i$ be the cell value at the $i^{th}$ row and $j^{th}$ column in the given clusters. For each pair of non-identical values $v_j^i$ and $v_k^i$ in the same cluster, except generating two candidate replacements $v_j^i \rightarrow v_k^i$ and $v_k^i \rightarrow v_j^i$, we also append the entry $(i, j)$ to $L[v_j^i \rightarrow v_k^i]$ and another entry $(i, k)$ to $L[v_k^i \rightarrow v_j^i]$.

**Updating Replacement Sets.** For each approved replacement lhs → rhs, if the users decide to replace lhs with rhs, for each entry $(i, j)$ in $L[\text{lhs} \rightarrow \text{rhs}]$, we replace the value $v_j^i$ (it must be lhs) with rhs. In addition, for each value $v_k^i$ within the same cluster as $v_j^i$, we update the replacement sets as follows. We remove the entry $(i, j)$ from $L[\text{lhs} \rightarrow v_k^i]$ and the entry $(i, k)$ from $L[v_k^i \rightarrow \text{lhs}]$, if $v_k^i$ is not identical to lhs. Moreover, we add the entry $(i, j)$ to $L[\text{rhs} \rightarrow v_k^i]$ and the entry $(i, k)$ to $L[v_k^i \rightarrow \text{rhs}]$, if $v_k^i$ is not identical to rhs. Note that, if a replacement set becomes empty in this process, which indicates the corresponding replacement no longer exists, we remove the corresponding replacement from $\Phi$. Since rhs must be an existing value in the give clusters, no new candidate replacements will be generated in this process. Similarly, in the case the users decide to replace rhs with lhs, we conduct the above process in the other way around.

## 7.2 Refine Groups by Structures

We observe that using current DSL, some replacements share a common transformation may look very different syntactically. In this case, it is hard for the users to make a single decision. To alleviate this problem, we propose to refine the groups by their *structures*. The candidate replacements in $\Phi$ are grouped together only if they share both the same transformation and the same structure.

In general, the structure of a replacement is acquired by uniquely mapping the two sides of the replacement to two sequences of pre-defined character classes (e.g., numeric and lowercase character classes). This is similar to the part-of-speech tagging in natural language processing [27].

Formally, the *structure* of a replacement $\varphi$, denoted by Struc($\varphi$), is based on decomposing each side of a replacement into a sequence of terms (character classes), drawn from the following:

- Regex-based terms:

  (1) *Digits*: $T_d = [0\text{-}9]+$     (2) *Lowercase letters*: $T_l = [\text{a-z}]+$
  (3) *Whitespaces*: $T_b = \backslash s+$    (4) *Capital letters*: $T_C = [\text{A-Z}]+$

- Single character terms:

  (5) The character cannot be expressed by regex-based terms, e.g., $T_-$ for the character '-'

Clearly, each character in any string will fall in *one and only one* of the above terms, such that the string can be uniquely represented. Next we show how to map a replacement to its structure. Initially, the structure of $s$, Struc($s$), is empty. We sequentially visit each character $s[i]$ in $s$ for $i \in [1, |s|]$. If $s[i]$ does not belong to any of the categories 1–4 above, i.e., $s[i]$ is a single character term, we append $T_{s[i]}$ to Struc($s$); otherwise, suppose $s[i]$ belongs to the category $x$ ($x \in [1, 4]$), we append $T_x$ ($T_d, T_l, T_C$ or $T_b$ depending on $x$) to Struc($s$) and skip all the consecutively subsequent characters in the same category. Finally, we obtain the structure Struc($s$) of $s$. For example, the structures of $s = 9$ and $t = 9$th are respectively Struc($s$) = $T_d$ and Struc($t$) = $T_d T_l$.

Definition 4 (Structure Equivalence). *Two replacements* $\varphi_1$ : lhs$_1$ → rhs$_1$ *and* $\varphi_2$ : lhs$_2$ → rhs$_2$ *are structurally equivalent, denoted by* Struc($\varphi_1$) ≡ Struc($\varphi_2$), *iff.* Struc(lhs$_1$) = Struc(lhs$_2$) *and* Struc(rhs$_1$) = Struc(rhs$_2$).

**Table 6: The dataset details**

|  | AuthorList | Address | JournalTitle |
|---|---|---|---|
| avg/min/max cluster size | 26.9/1/159 | 5.8/1/1196 | 1.8/1/203 |
| # of distinct value pairs | 51,538 | 80,451 | 81,350 |
| variant value pairs % | 26.5% | 18% | 74% |
| conflict value pairs % | 73.5% | 82% | 26% |

As it is less time consuming to get the structure of a replacement than calculating the pivot path, we first group the replacements in $\Phi$ by their structures. Specifically, for each replacement $\varphi$ in $\Phi$, we compute its structure $\textsc{Struc}(\varphi)$. All replacements in $\Phi$ are then partitioned into disjoint groups based on *structure equivalence*. For example, the two replacements $9 \rightarrow$ 9th and $3 \rightarrow$ 3rd will be grouped together, as they have the same structure $T_d \rightarrow T_d T_l$. By the end, we have a set of structure groups $\Phi_1, \Phi_2, \cdots$, which is a partition of $\Phi$. Then for each structure group $\Phi_i$, we invoke $\textsc{Unsupervised Grouping}(\Phi_i)$ as discussed in Section 5 to refine it into disjoint groups. To support the incremental grouping technique as discussed in Section 6, for each replacement $\varphi \in \Phi_i$, we use the structure group size $|\Phi_i|$ to initialize its upper bound. Then, whenever the first time a replacement in a structure group $\Phi_i$ is visited in Algorithm 7, we invoke $\textsc{Preprocessing}(\Phi_i)$ to build graphs and inverted index and recalculate tighter upper bounds. The rest remains the same.

### 7.3 Extending the DSL

The original DSL in [22] cannot describe the prefix and suffix (*a.k.a.* affix) relationship between the substrings in the input and output string. We extend the original DSL with two new string functions to support the affix relation. We discuss the details in Appendices D and F.

### 7.4 The Static Order of Functions

Similar to [22, 23, 34], we also force a static (partial) order for the position functions that locate the same position in the input string and the string functions that produce the same substring to improve the efficiency of our unsupervised method. We discuss the details in Appendix E.

## 8 EXPERIMENTS

The goal of the experiments is to evaluate the effectiveness and efficiency of our proposed unsupervised methods for standarizing variant values and golden record construction.

**Datasets.** We used the following three real-world datasets.

- AuthorList[4] contains information on 33,971 books. There are 1,265 clusters identified by matching their ISBN. Typical attributes include book name, author list, ISBN, and publisher. We used the author list attribute in the experiment, which contains 51,538 distinct value pairs. See more dataset details in Table 6.

- JournalTitle[5] contains 55,617 records concerning scientific journals. Attributes include journal title and ISSN. We clustered the journals by their ISSN numbers, resulting in 31,023 clusters. We used the journal title attribute in the experiment, which contains 80,451 distinct value pairs.

- Address[6] contains information on 17,497 applications for New York City Council Discretionary Funding. Attributes include the council member who started the application and the legal name, address, and Employer Identity Number (EIN) of the organization which applied for the funding. We clustered the applications by the EIN and used the address attribute in the experiment, which resulting in 3,038 clusters and 81,350 distinct value pairs.

**Setup.** We implemented our methods in C++, compiled using g++4.8, and did experiments on a server with Intel(R) Xeon(R) CPU E7-4830 @2.13GHz processors and 128 GB memory.

**Metrics.** To evaluate the efficiency of our unsupervised methods, we report the runtime for generating groups. For the effectiveness, we report the *precision*, *recall*, and *Matthews correlation coefficient* (MCC) of standarizing variant values.

**Table 7: Components for calculating the metrics**

|  | become identical | remain non-identical |
|---|---|---|
| variant value pairs | True Positives | False Negatives |
| conflict value pairs | False Positives | True Negatives |

Specifically, we first randomly sampled 1000 non-identical value pairs for each dataset and manually labeled each value pair as variant value pairs (i.e., they refer to the same value) or conflict value pairs (i.e., they refer to different values). We then ran our algorithm on the three datasets. After confirming a certain number of replacement groups generated by our methods and applying the approved ones to update the clusters, we checked the 1000 sample value pairs. Then, as shown in Table 7, true positives are the variant value pairs that become identical after updating, false negatives are the variant value pairs that remain non-identical after updating, false positives are the conflict value pairs that become identical after updating, and true negative are the conflict value pairs that remain non-identical after updating. We count the numbers of true positives (TP), false negatives (FN), false positives (FP), and true negatives (TN) and calculate the precision as $\frac{TP}{TP+FP}$, recall as $\frac{TP}{TP+FN}$, and

$$MCC = \frac{TP \times TN - FP \times FN}{\sqrt{(TP+FP)(TP+FN)(TN+FP)(TN+FN)}}.$$

MCC returns a value in $[-1, 1]$. The larger the better. We did not use the F1-score as the sizes of the positive class (variant value pairs) and negative class (conflict value pairs) were quite different, which could bias against the precision or recall [25]. The MCC is known to be a balanced metric even if the classes are of very different sizes [5].

---

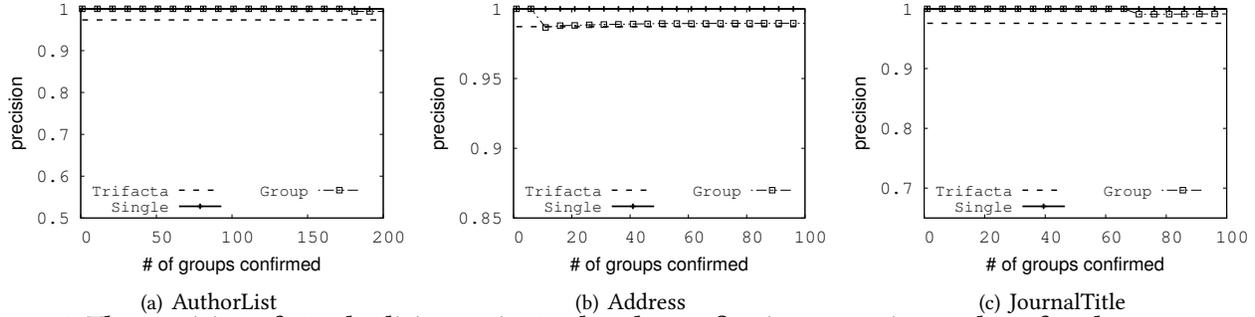

(a) AuthorList                              (b) Address                              (c) JournalTitle

**Figure 6: The precision of standardizing variant values by confirming a certain number of replacement groups**

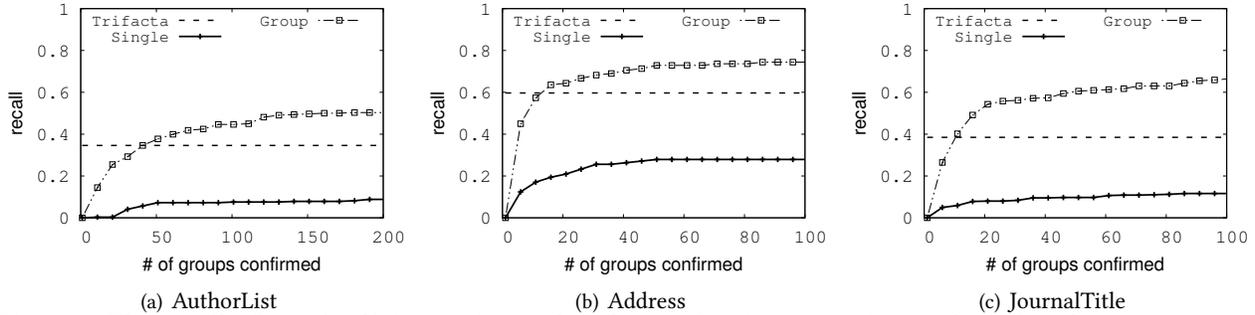

(a) AuthorList                              (b) Address                              (c) JournalTitle

**Figure 7: The recall of standardizing variant values by confirming a certain number of replacement groups**

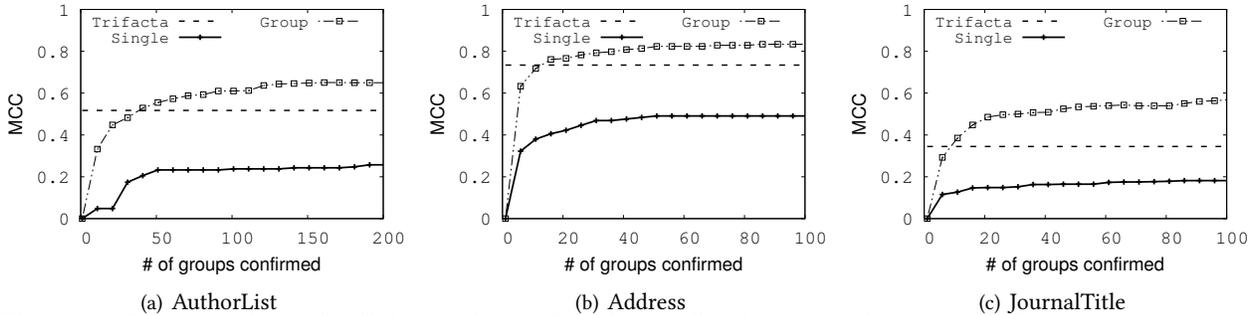

(a) AuthorList                              (b) Address                              (c) JournalTitle

**Figure 8: The MCC of standardizing variant values by confirming a certain number of replacement groups**

## 8.1 Effectiveness of Standardizing Data

We implemented two methods for standardizing variant values. (i) Single does not group the candidate replacements – each candidate replacement will be a group by itself. (ii) Group groups the candidate replacements in Φ by our proposed unsupervised methods: candidate replacements share the same pivot path (which corresponds to a transformation program) and structure will be grouped together.

We used Trifacta as our baseline method. Trifacta is a commercial data wrangling tool derived from DataWrangler [29]. It can apply some syntactic data transformations for data preparation, such as the regex-based replacing and substring extracting. Specifically, for each of the three datasets, we asked a skilled user to spend 1 hour on standardizing the dataset using Trifacta. Note that the user spent less than 20 minutes evaluating the groups in Single and Group in any of the experiments. Eventually, the user wrote 30-40 lines of wrangler code. For example, the following two lines of code were written to deal with groups C and E in Table 4.

<span style="color:red">REPLACE</span> *with*: '' *on*: '(\{any\}+)'
<span style="color:red">REPLACE</span> *with*: '$2 $3. $1' *on*: '(\{alpha\}+), (\{alpha\}+) (\{alpha\}.)'

The first rule removes all the contents between a pair of parentheses, including the parentheses themselves, such as "(edt)" and "(author)". The second rule changes the name formats. Note that many string transformation learning methods and tools have been proposed in recent years. However, all of them are semi-supervised and cannot be used or adapted for our problem. See more details in Section 9.

Figures 6, 7, and 8 show the results, where the *x*-axis represents the number of groups confirmed by a human and *y*-axis represents the corresponding precision, recall, and MCC of standardizing variant values as previously defined. The dotted lines are the results of the baseline Trifacta. With regard to recall, Group consistently achieved the best performance. Specifically, Group surpassed the recall of Trifacta and Single by up to 0.3 and 0.5 respectively. For example, on the JOURNALTITLE dataset, the recall of Group, Trifacta, and Single were respectively 0.66, 0.38, and 0.12. This is because, compared with the one-by-one verification in Single, the

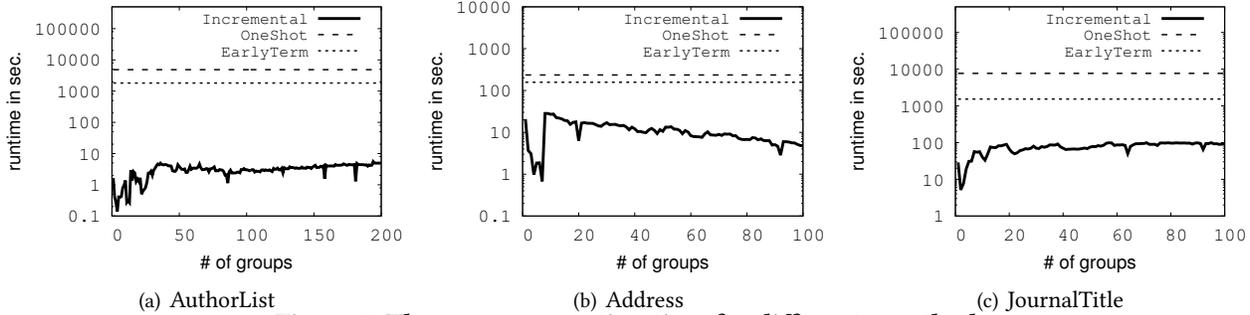

| (a) AuthorList | (b) Address | (c) JournalTitle |

**Figure 9: The group generation time for different menthods**

batch confirmation in Group is more effective in standardizing more data. For Trifacta, the users had to observe the data and write code. The code only covers a fraction of the data, whereas our unsupervised method judiciously presents the most frequent and 'profitable' groups for the user to verify.

All the methods achieved very high precision as they all had a human in the loop. Specifically, Single achieved 100% precision, while Group and Trifacta achieved precision above 99% and 97%. This is because the one-by-one checking of Single is more fine-grained than the batch verification of Group, while Trifacta applied the code globally, which may introduce some errors. Nevertheless, the batch verification in Group and the human-written code in Trifacta were very effective with regard to precision. Overall, Group achieved the best MCC. It outperformed Trifacta and Single by up to 0.2 and 0.4 respectively. For example, on JournalTitle, the MCC of Group, Trifacta, and Single was respectively 0.57, 0.34, and 0.18, for the same reasons as discussed above.

Note that for all three datasets, the user spent less than 20 minutes in confirming the groups. Though Single took a little less human time than Group, its performance was much worse than that of Group as discussed above. In total, the user approved 70, 39, and 22 groups in Group out of the 200, 100, and 100 groups presented in AuthorList, Address, and JournalTitle. The denied groups were mostly because of the logic. For example, one group in AuthorList transposes the authors' order and thus got denied.

### 8.2 Efficiency of the Grouping Algorithms

In this section, we evaluate the efficiency of our grouping methods. We implemented three methods. (i) OneShot uses the vanilla UnsupervisedGrouping method to generate groups as discussed in Section 5.1. (ii) EarlyTerm improves OneShot by the two early termination techniques as discussed in Section 5.2. (iii) Incremental uses our incremental grouping method to generate groups as discussed in Section 6. We reported the group generation time for these methods. Figure 9 shows the results. In the figure, the two dotted lines for OneShot and EarlyTerm show their upfront costs. The solid line for Incremental gives the runtime of GenerateNextLargestGroup at each invocation.

We can see from the figure that Incremental achieved the best performance. It improved the upfront cost of EarlyTerm by up to 3 orders of magnitude, while EarlyTerm outperformed OneShot by 2-10 times. For example, for the AuthorList dataset, the upfront cost for OneShot, EarlyTerm, and Incremental were respectively 4900 seconds, 1800 seconds, and 1.6 seconds. This is because the two optimizations in EarlyTerm can avoid a lot of unnecessary invocations of SearchPivot in finding the pivot path compared to OneShot. In addition, Incremental only generates the largest group at each time and thus can skip many unnecessary candidate replacements in $\Phi$. Note that all these three methods had the same effectiveness for standardizing variant values as they are guaranteed to produce the same groups.

Note that we limited the maximum path length to 6 when searching for the pivot path in all the experiments in order to have the programs finish in reasonable time. Actually there is a trade-off between the effectiveness and efficiency of group generation. If it takes too much time to generate the groups, we can limit the maximum path length or use sampling to accelerate group generation. See more details in Appendix E.

### 8.3 Improvement on Entity Consolidation

In this section, we evaluate the effectiveness of our algorithm in assisting truth discovery. For this purpose, we collected ground truth for 100 random clusters for each dataset. For AuthorList, we used the same manually created ground truth as the previous work [15]. For JournalTitle and Address, we manually searched for the ISSN in www.issn.cc and the EIN in www.guidestar.org to create the ground truth for each cluster. We used the dataset without any normalization except converting all characters to lowercase.

We first used the majority consensus (MC) to generate the golden values for each cluster and then compared the golden values with the ground truth. If they refer to the same entity, we increase TP (true positive) by 1; otherwise, we increase FP (false negative) by 1. Note that if there are two values with the same frequency, MC could not produce a golden value. Next we processed the original dataset with our algorithm and re-ran MC to create the golden values. We reported the precision before and after using our techniques. Table 8 shows the results. We observe that our method indeed helped improve the precision of MC. In particular, on JournalTitle, MC produced a precision of 33.5% before using our algorithm. After processing JournalTitle with our algorithm, MC produced a precision of 84%, which is an improvement of over 40%. This is attributed to our effective variant value standardizing method, which correctly consolidated most of the duplicate values. On the other datasets, the improvement was less dramatic but still significant.

**Table 8: Precision improvement for MC**

|        | AuthorList | Address | JournalTitle |
|--------|------------|---------|--------------|
| before | .51        | .32     | .335         |
| after  | .65        | .47     | .840         |

## 9 RELATED WORK

Our work is related but orthogonal to work in string transformation, entity resolution, entity consolidation, truth discovery, and data fusion. We review them in this section.

**String Transformations.** Significant research has been conducted to transform string values. All of them are semi-supervised approaches, i.e., they need user-provided examples. Moreover, they are limited in learning the string transformation from homogeneous data, one at a time. In contrast, our string transformation learning method is unsupervised. Our approach generates a large number of potentially dirty examples (candidate replacements) from the heterogeneous data and learns the string transformations (replacement groups) all at once. As the input data in entity consolidation usually come from different sources with different formats and thus are heterogeneous, the semi-supervised methods cannot be used or adapted for entity consolidation.

Specifically, FlashFill [22] and BlinkFill [34] proposed to use program synthesis techniques [37] to learn a consistent transformation in a pre-defined DSL from a few user-provided input-output examples. Recently, a couple of work [14, 28] propose to use neural network to guide the program search in FlashFill and BlinkFill. DataXFormer [2, 3] proposes to search string transformations from web tables, webforms, and knowledge bases based on the user-provided examples. Similarly, He et al. [24] developed a search engine to find string transformations that are consistent to the user-provided examples from large scale codebases. Arasu et al. propose to learn string transformation rules from given examples [4]. Singh et al. present an approach to learn semantic string transformations from user provided examples [35]. DataWrangler [29] and its commercial descendant Trifacta has some string transformation functionality such as regex-based replacing, string splitting, extracting substrings, etc. It also provides layout transformations for structured data, similar to the Foofah system [26]. Wang et al. present a probabilistic approach to learn string transformations for spelling error correction and web search query reformulation [41].

**Entity Consolidation.** Entity consolidation aims at merging duplicate records [8, 16]. Entity consolidation is typically user-driven. For example, Swoosh [6] provides a unified interface that relies on the users to define the Merge function to specify how to merge two duplicate records. The conventional wisdom for entity consolidation is to use a Master Data Management (MDM) product [1]. MDM systems include a match-merge component, which is based on a collection of human-written rules. However, it is well understood that MDM solutions do not scale to complex problems, especially ones with large numbers of clusters and records.

**Truth Discovery and Data Fusion.** Truth discovery and data fusion [15, 17, 31, 33, 39, 44] can be used for entity consolidation. Given a set of claims made by multiple sources on a specific attribute, truth discovery and data fusion decide whether each claimed value is true or false and compute the reliability of each source. Solutions to these problems include models that use prior knowledge about the claims or source reputation [44], methods that consider the trustworthiness of a source as a function of the belief in its claims and the belief score of each claim as a function of the trustworthiness of the corresponding sources [31, 33], methods that consider other aspects such as source dependencies and truth evolution [15, 17]. In addition, there are approaches that try to resolve data conflicts by optimizing data quality criteria, such as data currency [19] and data accuracy [9], which selects the most recent value and the most accurate value, respectively. These works solve a different problem. They can be used to compute golden records. However, standardizing the variant values to the same canonical format using our method before applying them can improve their performance.

**Entity Resolution.** Entity resolution deals with the problem of identifying records that represent the same real-world entity [18]. Generally speaking, there are machine learning-based [20] and rule-based [40, 42, 43] solutions for entity resolution. Machine learning-based methods include SVM-based [7], decision tree-based [21, 38], clustering-based [10], and Markov logic-based [36] methods. Entity resolution techniques are orthogonal to our contributions, since we take their output, i.e., clusters of duplicate records, as our input and identify golden records within those clusters.

## 10 CONCLUSIONS

In this paper, we proposed an unsupervised string transformation learning method for entity consolidation. Instead of directly applying existing solutions for entity consolidation, we first standardize the variant data. Specifically, we first enumerate the attribute value pairs in the same cluster. Then, we employ an unsupervised method to group value pairs that can be transformed in the same way. Finally we confirm the groups with a human and apply transformations in the approved groups to standardize the variant data. Experiments on real-world datasets show that our solution can effectively standardize variant values and significantly improve the performance versus a state of the art data wrangling tool.

# A    ADDITIONAL FINE-GRAINED CANDIDATE REPLACEMENTS

Another way to generate additional fine-grained candidate replacements is to enumerate every pair of non-identical values in the same cluster, split them into segments, and align these segments to compose candidate replacements.

For illustration purpose, we split the values by whitespaces and leverage the longest common subsequence (LCS) method to obtain the alignment. More specifically, for each pair of values within the same cluster in $\mathbf{C}^i$, we first split them into token sequences by whitespaces and then calculate their LCS. The LCS naturally aligns the two token sequences. Each aligned pair of non-identical subsequences composes a pair of candidate replacements. A similar approach is studied in [41], but at the character level.

*Example A.1.* Consider two attribute values $r_1[\texttt{Address}] =$"9 St, 02141 Wisconsin" and $r_2[\texttt{Address}] =$"9th St, 02141 WI". Their LCS is "St, 02141". The two aligned non-identical subsequences will produce four candidate replacements: 9 → 9th, 9th → 9, Wisconsin → WI, and WI → Wisconsin as highlighted in the figure below.

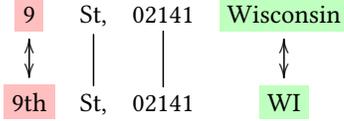

There are also other ways to align the token sequences and generate candidate replacements [4]. For example, we can use the alignment (the consecutive matching operations) calculated from the Damerau-Levenshtein distance [11] to obtain the candidate replacements. Our proposed technique is independent from the candidate replacement generation methods and can work with any one or a combination thereof.

## B  THE DOMAIN-SPECIFIC LANGUAGE

This section briefly summarizes the domain-specific language designed by Gulwani [22] in a formal way. The readers are encouraged to read the original paper for a comprehensive understanding of the DSL.

In the following, we use $\mathbf{s}[i, j)$ to denote the substring $\mathbf{s}[i, j - 1]$ and define the beginning and ending positions of $\mathbf{s}[i, j)$ as $i$ and $j$, denoted by $\text{beg}(\mathbf{s}[i, j)) = i$ and $\text{end}(\mathbf{s}[i, j)) = j$, respectively. For example, consider the string "Mary", we have $\mathbf{s}[1, 2) = \mathbf{s}[1, 1] = \texttt{M}, \text{beg}(\mathbf{s}[1, 2)) = 1$, and $\text{end}(\mathbf{s}[1, 2))=2$.

The DSL defines *position function* and *string function*, which all apply to the input string $\mathbf{s}$ (*a.k.a. global parameter*). A position function returns an integer that indicates a position in $\mathbf{s}$. It can help locate the substrings of $\mathbf{s}$ produced by the string functions. Next we give the formal definitions.

**Position Functions.** There are two *position functions*.

- $\texttt{ConstPos}(k)$: given an integer $k$, it outputs:
  ○ $k$, *if* $0 < k \leq |\mathbf{s}| + 1$, i.e., a forward search, or
  ○ $|\mathbf{s}|+2+k$, *if* $-(|\mathbf{s}| + 1) \leq k < 0$, i.e., a backward search.
- $\texttt{MatchPos}(\tau, k, \texttt{Dir})$: this function can express semantics like "*the ending position of the $k^{th}$ digital substring in $\mathbf{s}$*". Formally, it takes a term $\tau$ in $\{\texttt{T}_C, \texttt{T}_d, \texttt{T}_l, \texttt{T}_b\}$, an integer $k$, and a binary-state variable $\texttt{Dir}$ ($\texttt{B}$ for the beginning position and $\texttt{E}$ for the ending position) as input. Let $\mathbf{s}_{(\tau, i)}$ be the $i^{th}$ matched substring of the term $\tau$ in $\mathbf{s}$ and $m$ be the total number of matches of $\tau$ in $\mathbf{s}$. The function returns one of the following positions:
  ○ $\text{beg}(\mathbf{s}_{(\tau, k)})$, *if* ($0 < k \leq m$ and $\texttt{Dir} = \texttt{B}$),
  ○ $\text{end}(\mathbf{s}_{(\tau, k)})$, *if* ($0 < k \leq m$ and $\texttt{Dir} = \texttt{E}$),
  ○ $\text{beg}(\mathbf{s}_{(\tau, m+1+k)})$, *if* ($-m \leq k < 0$ and $\texttt{Dir} = \texttt{B}$), or
  ○ $\text{end}(\mathbf{s}_{(\tau, m+1+k)})$, *if* ($-m \leq k < 0$ and $\texttt{Dir} = \texttt{E}$)

*Example B.1.* Consider $\mathbf{s} = $"Lee, Mary", where $|\mathbf{s}| = 9$.
- $\texttt{ConstPos}(2) = 2$ and $\texttt{ConstPos}(-5) = 9 + 2 - 5 = 6$.
- $\texttt{MatchPos}(\texttt{T}_C, 2, \texttt{B}) = \text{beg}(\mathbf{s}_{(\texttt{T}_C, 2)}) = \text{beg}(\mathbf{s}[6, 7)) = 6$ and $\texttt{MatchPos}(\texttt{T}_C, 2, \texttt{E}) = \text{end}(\mathbf{s}_{(\texttt{T}_C, 2)}) = \text{end}(\mathbf{s}[6, 7)) = 7$. They are the beginning and ending positions of the $2^{nd}$ match $\mathbf{s}_{(\texttt{T}_C, 2)} = \mathbf{s}[6, 7) = $"M" in $\mathbf{s}$ of the capital letter term $\texttt{T}_C$.

Note that besides the 4 regex-based terms as defined in Section 7.2, we also allow using the *constant string terms* in MatchPos function. A constant string term $\texttt{T}_{\texttt{str}}$ matches and only matches the constant string $\texttt{str}$. Since the single-character term is subsumed by the constant string term, we do not use it in the MatchPos function.

**String Functions.** There are also two *string functions*.
- $\texttt{ConstantStr}(\mathbf{x})$ simply outputs the input string $\mathbf{x}$.
- $\texttt{SubStr}(l, r)$ outputs the substring $\mathbf{s}[l, r)$ given two input integers $l$ and $r$, where $l < r$. Note that position functions can be taken as parameters in this function.

*Example B.2.* Consider $\mathbf{s} = $"Lee, Mary", we have
- $\texttt{ConstantStr}(\texttt{MIT}) = \texttt{MIT}$,
- $\texttt{SubStr}(\texttt{MatchPos}(\texttt{T}_C, 1, \texttt{B}), \texttt{MatchPos}(\texttt{T}_l, 1, \texttt{E})) = $"Lee".

**Definition 5 (Transformation Program).** *A transformation program is:*

$$\rho(\mathbf{s}) := f_1 \oplus f_2 \oplus \cdots \oplus f_n$$

*where $\mathbf{s}$ is the input, "$\oplus$" is a string concatenation operator, and $f_i$ is a string function. Given an input string, the transformation program returns the concatenation of the outputs of all its string functions.*

*Example B.3.* Consider the transformation program $f_1 \oplus f_2 \oplus f_3$, where

- $f_2 : \texttt{SubStr}(\texttt{MatchPos}(\texttt{T}_C, 1, \texttt{B}), \texttt{MatchPos}(\texttt{T}_l, 1, \texttt{E}))$
- $f_3 : \texttt{SubStr}(\texttt{MatchPos}(\texttt{T}_b, 2, \texttt{B}), \texttt{MatchPos}(\texttt{T}_C, -1, \texttt{E}))$
- $f_1 : \texttt{ConstantStr}(.\ )$

Given the input string $\mathbf{s} = $"Lee, Mary", the program produces an output string $\mathbf{t} = $"M. Lee" as follows:

$$\underbrace{\texttt{M}}_{f_1} \oplus \underbrace{.}_{f_2} \oplus \underbrace{\texttt{Lee}}_{f_3} = \underbrace{\texttt{M. Lee}}_{\textbf{Program}}$$

**Algorithm 8:** BuildTransformationGraph($\mathbf{s}$, $\mathbf{t}$)

---

**Input:** $\mathbf{s}$: the left hand string; $\mathbf{t}$: the right hand string.

**Output:** G(N, E): the transformation graph for $\mathbf{s} \rightarrow \mathbf{t}$.

1 **begin**
2    let $\mathcal{P}$ be a two-dimensional array;
3    **foreach** $k^{th}$ *match of the term* $\tau$, $\mathbf{s}[x, y)$ **do**
4      let $m_\tau$ be the number of total matches of $\tau$ in $\mathbf{s}$;
5      $\mathcal{P}[x] \leftarrow$ MatchPos($\tau, k$, B);
6      $\mathcal{P}[x] \leftarrow$ MatchPos($\tau, k - m_\tau - 1$, B);
7      $\mathcal{P}[y] \leftarrow$ MatchPos($\tau, k$, E);
8      $\mathcal{P}[y] \leftarrow$ MatchPos($\tau, k - m_\tau - 1$, E);
9    **foreach** $1 \leq k \leq |\mathbf{s}| + 1$ **do**
10      $\mathcal{P}[k] \leftarrow$ ConstPos($k$);
11      $\mathcal{P}[k] \leftarrow$ ConstPos($k - |\mathbf{s}| - 2$);
12    N $= \{n_1, \cdots, n_{|\mathbf{t}|}\}$;
13    **foreach** $1 \leq i < j \leq |\mathbf{t}|$ **do**
14      add the edge $e_{i,j}$ from $n_i$ to $n_j$ to E;
15      add the label ConstantStr($\mathbf{t}[i, j)$) to $e_{i,j}$;
16      **foreach** $x$ *and* $y$ *s.t.* $\mathbf{s}[x, y) = \mathbf{t}[i, j)$ **do**
17        **foreach** $f \in \mathcal{P}[x]$ *and* $g \in \mathcal{P}[y]$ **do**
18          add the label SubStr($f, g$) to $e_{i,j}$;
19    **return** G $= (N, E)$
20 **end**

## C BUILD A TRANSFORMATION GRAPH

Algorithm 4 gives the pseudo-code for constructing a transformation graph for a replacement $\mathbf{s} \rightarrow \mathbf{t}$. It first builds a two-dimensional array $\mathcal{P}$ to keep all the possible position functions to represent a position in $\mathbf{s}$. To this end, it finds all the matches of any term $\tau$ and any substring $\mathbf{s}[x, y)$ of $\mathbf{s}$. For each of these matches, it respectively adds the beginning and ending positions of the match to $\mathcal{P}[x]$ and $\mathcal{P}[y]$ (Lines 3 to 8). In addition, for each $1 \leq k \leq |\mathbf{s}| + 1$, it adds the constant positions ConstPos($k$) and ConstPos($k - |\mathbf{s}| - 2$) to $\mathcal{P}[k]$ (Lines 9 to 11). Next, it constructs a set N of $|\mathbf{t}| + 1$ nodes and build an edge $e_{i,j}$ for each substring $\mathbf{t}[i, j)$ of $\mathbf{t}$. Afterwards, for each substring $\mathbf{t}[i, j)$, it adds a constant string label ConstantStr($\mathbf{t}[i, j)$) to the edge $e_{i,j}$ (Line 15). In addition, for each substring $\mathbf{s}[x, y)$ that matches $\mathbf{t}[i, j)$, it adds all the labels SubStr($f, g$) to $e_{i,j}$ where $f$ and $g$ are position functions in $\mathcal{P}[x]$ and $\mathcal{P}[y]$ (Line 18). Finally it returns the transformation graph (Line 19).

**Complexity Analysis.** Given a replacement $\mathbf{s} \rightarrow \mathbf{t}$, the complexity of this algorithm is $O(|\mathbf{s}|^2 |\mathbf{t}|^2)$. We need to identify all the matches between any term and any substring of $\mathbf{s}$ to build the position array $\mathcal{P}$. To find all the matches of a regex-based term, we need to sequentially scan $\mathbf{s}$. Note that there are a constant number of regex-based terms (in our case it is 4). To find all the matches of the constant-based terms, we need to enumerate every substring of $\mathbf{s}$. Thus constructing the position array $\mathcal{P}$ needs $O(|\mathbf{s}|^2)$ time. For each substring $\mathbf{t}[i, j)$, we build an edge $e_{i,j}$. There are up to $|\mathbf{s}|$ substrings of $\mathbf{s}$ that can match $\mathbf{t}[i, j)$. For each matched substring, there are $O(|\mathbf{s}|^2)$ string functions. Thus the total time complexity is $O(|\mathbf{s}|^2 |\mathbf{t}|^2)$.

## D AFFIX STRING FUNCTIONS

We extend the DSL to support the affix semantics in this section. The current DSL cannot express the affix semantics. For example, using the language, there is no program that is consistent with the two replacements Street $\rightarrow$ St and Avenue $\rightarrow$ Ave[7]. This is because FlashFill [22, 23] uses the program to calculate an output string from an input string and the program and the string functions (SubStr and ConstantStr) are required to return a single deterministic output string given a specific input string. However, one input string may have multiple affixes (suffixes and prefixes) and thus current DSL [22] cannot support affix semantics. In general, our program can accept any string function that takes an input string and returns one or multiple output strings. For a specific input string lhs, if rhs is one of the output strings of a program, the program can describe how lhs is transformed to rhs and further be used to group replacements.

Consequently, we design two string functions Prefix($\tau, k$) and Suffix($\tau, k$) to be used as labels on the edges in the transformation graph as defined below.

**Definition 6 (Affix Labels).** *For each edge $e_{i,j}$ in the transformation graph of $\mathbf{s} \rightarrow \mathbf{t}$, we add the following two kinds of labels to it:*

- Prefix($\tau, k$): *if* $\mathbf{t}[i, j)$ *is a prefix of the $k^{th}$ match of the regex-based term $\tau$ in $\mathbf{s}$.*
- Suffix($\tau, k$): *if* $\mathbf{t}[i, j)$ *is a suffix of the $k^{th}$ match of the regex-based term $\tau$ in $\mathbf{s}$.*

*Example D.1.* Consider the transformation graph of the replacement Street $\rightarrow$ St, the edge $e_{2,3}$ has a label Prefix($\mathrm{T}_l$, 1) as 't' is a prefix of the $1^{st}$ match of the lower case term 'treet'. Similarly, the graph of Avenue $\rightarrow$ Ave also has an edge $e_{2,4}$ containing the label Prefix($\mathrm{T}_l$, 1) as 've' is a prefix of the $1^{st}$ match of the lower case term 'venue'. A consistent program for both Street $\rightarrow$ St and Avenue $\rightarrow$ Ave is SubStr(MatchPos($\mathrm{T}_C$, 1, B), MatchPos($\mathrm{T}_C$, 1, E))⊕Prefix($\mathrm{T}_l$, 1).

The pivot path search algorithm and the incremental grouping algorithm work all the same as before and with the two additional kinds of labels. Interestingly, adding the affix labels not only makes the DSL more expressive but also largely improves the efficiency of our unsupervised methods. This is because with the additional affix labels, our algorithm can reach a transformation path earlier in SearchPivot, which will set the local and global thresholds to enable early termination. The group D in Table 4 share a transformation path involving both `Prefix` and `Suffix` string function.

---

[7]try it here: https://rise4fun.com/QuickCode/1p

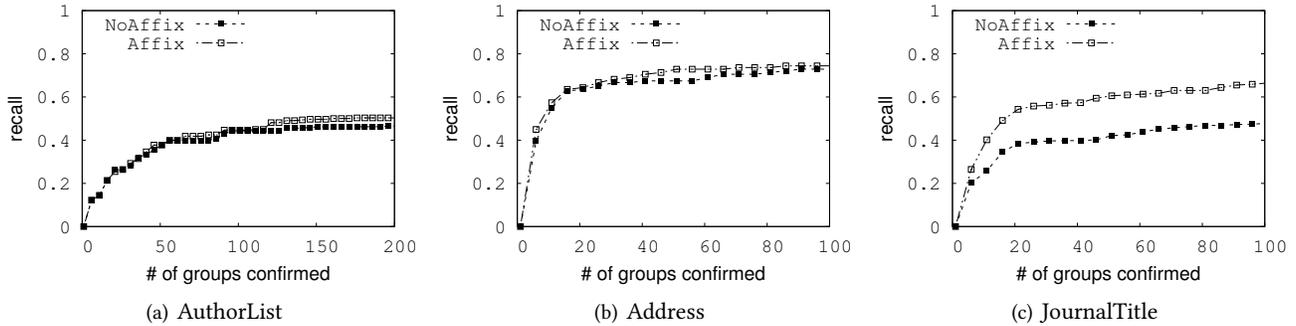

**Figure 10: The recall of standardizing variant values with and without the affix string functions**

## E THE STATIC ORDER OF FUNCTIONS

In this section, we design a static partial order for the position/string functions. Then we skip a position (string) function $f$ if there is a larger one in the static order that locates the same position in $\mathbf{s}$ (produce the same string) as $f$. Next we discuss the static orders.

**The Static Order of Position Functions.** Similarly to [22, 34], we also use a static order for the position functions that locate the same position in $\mathbf{s}$ to improve the efficiency. In this order, the regex-based term with wider character class is better than the narrower ones.

**The Static Order of Affix String Functions.** We only add the affix string function labels for the longest prefixes and suffixes to the transformation graph. For example, if both $\mathbf{t}[i, j]$ and $\mathbf{t}[i, j + 1]$ are prefixes of a match $\mathbf{s}[x, y]$ of a predefined regex $\tau$, we do not add the label `Prefix`$(\tau, k)$ to $e_{i,j}$ as the substring $\mathbf{t}[i, j]$ is shorter than $\mathbf{t}[i, j + 1]$.

**The Static Order of Constant Strings.** To reduce the search space of the pivot path, we propose a heuristic to score the constant strings. Then for each position in $\mathbf{s}$, we only use the constant string term with highest score to located it. In addition, we add `ConstantStr`$(\mathbf{t}[i, j))$ to the label set of the edge $e_{i,j}$ only if none of $\mathbf{t}[k, i)$ or $\mathbf{t}[j, l)$ for any $1 \le k < i$ and $j < l \le |\mathbf{t}| + 1$ has a larger score than $\mathbf{t}[i, j)$. Next we discuss the scoring scheme.

We observe that the constant string functions used in our pivot path are usually frequently appeared in the dataset. For example, the `ConstantStr(Mr.)` in "Mr. Lee" and "Mr. Smith". Thus we use the string frequency as an indicator.

However, the the single-character strings always have the largest frequency. To alleviate this problem, we use the string frequency within a structure group as another indicator. In particular, suppose the frequency of a constant string term $\tau$ within a structure group and within the dataset is $\text{freq}_{\text{Struc}}(\tau)$ and $\text{freq}_{\text{Global}}(\tau)$, we use the score $\frac{\text{freq}_{\text{Struc}}(\tau)}{\sqrt{\text{freq}_{\text{Global}}(\tau)}}$ to rank constant string terms in this structure group. This would prefer constant string term appears frequently within its structure group but infrequent outside the group. Note that the single-character strings are frequent both inside and outside a group and are less preferred in this ranking scheme. We square the frequency in the dataset to strengthen the impact of the frequency in groups.

Note that, when there are a huge number of replacements in $\Phi$ such that our unsupervised methods are not efficient enough, we can randomly sample a small part of replacements in $\Phi$ and use the transformation path contained by the largest number of the samples as the pivot path. This will improve the efficiency of our pivot path selection method as the inverted list intersection takes less time.

We can also set a maximum path length $\theta$ and skip invoking the recursive algorithm (Line 11 in Algorithm 3) if the maintained path $\rho'$ is longer than $\theta$.

## F EVALUATING AFFIX STRING FUNCTIONS

In this section, we evaluate the effectiveness of the two affix functions, `Suffix` and `Prefix`. We implemented two methods Affix and NoAffix, one with and the other without the two string functions in the transformation graph. We varied the number of confirmed groups and reported the recall of the two methods. Figure 10 gives the results. We observed that Affix always produced higher recall than NoAffix. This is because some replacements cannot be grouped together without the two affix string functions. The precision were all close to 100% and the MCC result was similar to that of the recall.